\documentclass[twocolumn,showpacs,preprintnumbers,superscriptaddress,10pt,aps,prb]{revtex4-2}
\usepackage{amsmath,amssymb}
\usepackage{graphicx}
\usepackage{physics}
\usepackage{amsthm}
\usepackage[dvipsnames]{xcolor}
\usepackage[colorlinks=true,linktoc=page,linkcolor=magenta,citecolor=magenta]{hyperref}

\newcommand\crule[3][black]{\textcolor{#1}{\rule{#2}{#3}}}

\begin{document}
 \title{Broken symmetry and competing orders in Weyl semimetal interfaces}

\author{Ritajit Kundu}
\affiliation{Department of Physics, Indian Institute of Technology Kanpur, Kanpur 208016, India}
\author{H.A. Fertig}
\affiliation{Department of Physics, Indiana University, Bloomington, IN 47405}
\affiliation{Quantum Science and Engineering Center, Indiana University, Bloomington, IN, 47408}
\author{Arijit Kundu}
\affiliation{Department of Physics, Indian Institute of Technology Kanpur, Kanpur 208016, India}
	
\begin{abstract}
We consider interaction-induced broken symmetry states of two Weyl semimetal surfaces with multiple Fermi-arc (FA) states. In the presence of inter- and intra-surface Coulomb interactions, multiple broken symmetries may emerge which coexist and/or compete with one another.  Interlayer exciton condensates involving different FA flavors are shown to form, with amplitudes determined by the strength of interactions and the degree of nesting among the arcs.  For FA pairs which are well-separated in momentum with strong nesting, the resulting state is a particle-hole analog of a Fulde–Ferrell–Larkin–Ovchinnikov (FFLO) superconductor.  Intralayer interactions moreover induce charge density wave (CDW) ordering, so that the most general state of the system is a supersolid.  These orderings in principle carry signatures in non-linear behavior and narrow band noise in Coulomb drag transport measurements.
\end{abstract}

\color{black}
\maketitle

\section{Introduction}
Weyl semimetals (WSMs) are three dimensional topological systems with an even number of band-touching points (Weyl nodes) in their bulk band-structure~\cite{WSMreview,WSMBook}.
Because of their intrinsic topology,
non-overlapping surface projections of Weyl nodes connect endpoints of disjoint Fermi surface sections known as Fermi-arcs (FAs). FAs host surface states that disperse in a quasi-one-dimensional manner. There are extensive ongoing efforts to identify material candidates for WSMs, both theoretically and experimentally. Examples of such materials include TaAs~\cite{taas}, NbAs~\cite{nbas} and, more recently, CoSi, Co$_3$Sn$_2$S$_2$, for which FA modes have been identified in ARPES and quasiparticle interference experiments~\cite{CoSi,CoSnS1,CoSnS2}. Although they may lack topological protections, FAs of  Dirac semimetals, such as Na$_3$Bi and Cd$_3$As$_2$,  have been also identified in recent times~\cite{FADSM01,FADSM02,FADSM1,FADSM2,FADSM3,FADSM4}.

\begin{figure}[t]
	\centering
	\includegraphics[width=.44\textwidth]{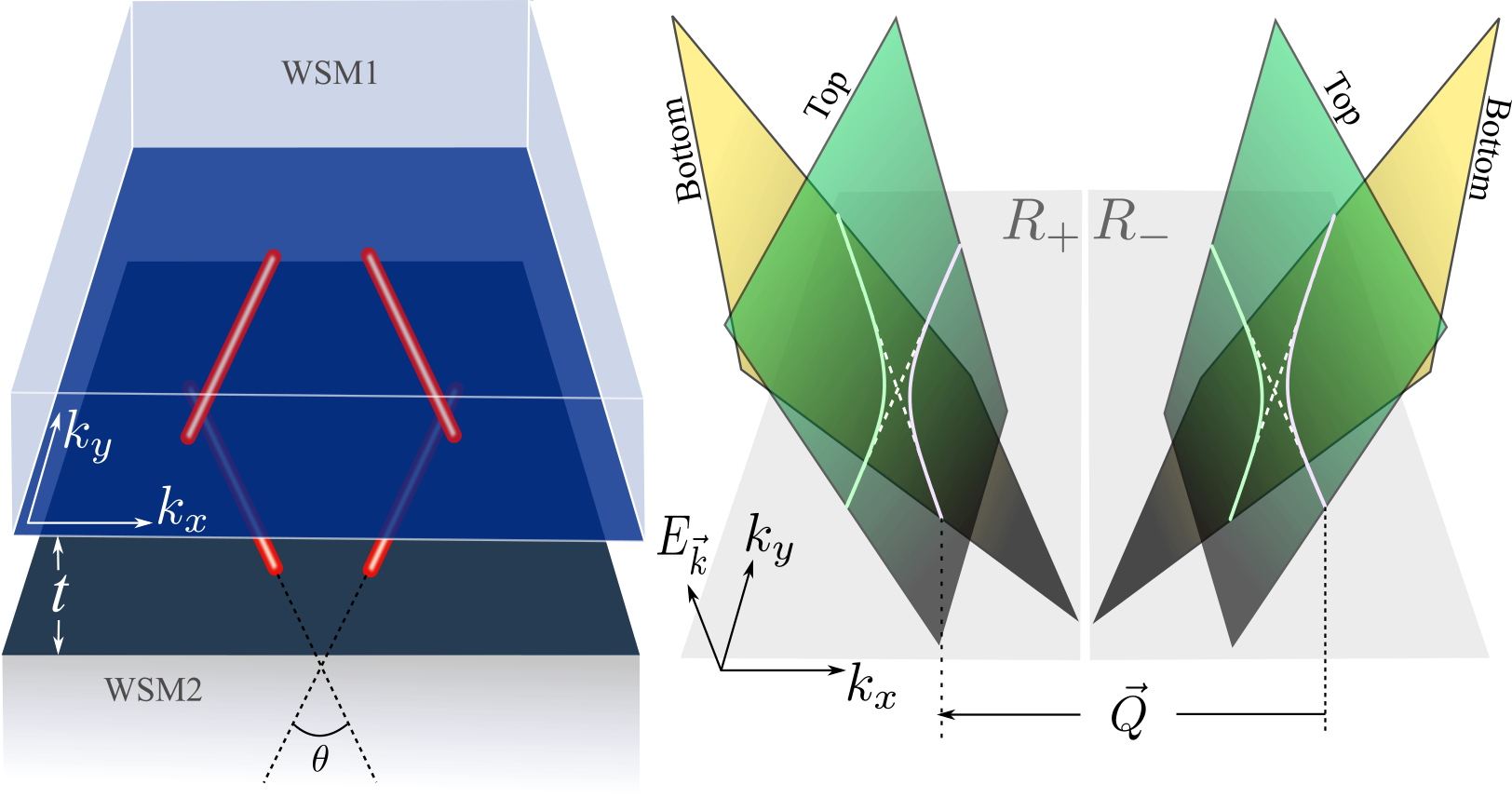}
	\caption{Left: Two WSM surfaces with FAs indicated in the surface Brillouin zones, which are separated by a dielectric slab of thickness $t$. Two FAs reside on each surface with an angle $\theta$ between them. Right: Dispersions of the FAs. White dashed lines indicate Fermi surfaces without interactions. With interactions the Fermi surfaces distort to the solid white lines, allowing CDW and FFLO order to form with nesting vector $\vec{Q}$.}\label{fig:setup}
\end{figure}

Interactions may introduce interesting physics in WSMs, involving either or both the bulk states and the FA states.  For example, collective excitations confined to the surfaces are expected to be supported ~\cite{weylplsurf1,weylplsurf2,weylplsurf3,weylplsurf4,weylplsurf5,weylplsurf6,weylplsurf7,weylplsurf8,weylplsurf9,weylplsurf10}, as are bulk excitonic modes and density-wave instabilities~\cite{WSMexcitonbulk1,WSMexcitonbulk2}, among other possibilities~\cite{WSMBook}. Interesting effects also occur when two Weyl systems are brought together.  For example, intricate reconstruction of FA geometry can sometimes occur due to inter-surface tunneling \cite{Dwivedi_2018,Murthy_2020,Abdulla_2021}.  In the absence of tunneling, inter-surface Coulomb interactions may induce coherent particle-hole processes involving FA states of both surfaces, leading to new collective excitations and broken symmetry states.  This is the subject of our study.
As explained below, we find that a number of symmetries may break in such systems: the local gauge symmetry, which conserves particle number of each layer, due to inter-surface exciton condensation ~\cite{excitonWSM}, in similarity with other bi-layer systems, such as in graphene~\cite{excitongraphene1,excitongraphene2,excitongraphene3}; translational symmetry, through the formation of charge-density-wave (CDW) order; and, in each case, coherences may set in among different pairs of arcs on the same or different surfaces, yielding multiple ways in which these kinds of orders set in.  As we shall see, while these orderings coexist, they also compete, leading to quantum phase transitions among different realizations of the broken symmetries with variations of the system parameters.

Associated with these broken symmetry states are Goldstone modes. The broken translational symmetry characteristic of CDW order leads to phonon modes, which at zero wavevector becomes a sliding mode that is generically pinned by disorder \cite{Gruner_1988}.  Exciton condensation yields gapless superfluid modes \cite{Snoke_2002,excitongraphene4,Fogler_2014,Combescot_2017} which in such double layer systems is realized as a dissipationless counterflow current.  Moreover, very weak tunneling between surfaces may yield Josephson-like transport behavior between them \cite{Eisenstein_2014,josexciton}.
Such collective behavior can be observed in a variety of transport experiments~\cite{excitonsup1,excitonsup2,cdrag,Eisenstein_2014,cdrag2}.

As a paradigm of such systems, we consider a setup of two capacitively coupled WSM surfaces, each hosting FAs (see Fig.~\ref{fig:setup}).  For simplicity we consider straight arcs, although our qualitative results do not depend significantly on this simplification (see Appendix \ref{app:construction_of_2_pair_of_tilted_fermi_arcs}).  We find that electron-hole coherence may develop among some or all of the arcs, depending on their relative angles and interaction strengths.  For arcs on different surfaces  with common in-plane wavevectors, strong interlayer coherence can develop at these common wavevectors \cite{excitonWSM}.  Interlayer coherence can also develop between arcs on different surfaces whose wavevectors are remote from one another, and surprisingly these coherences can be stronger than the direct case, particularly when the the arcs are nested. Such finite momentum interlayer ordering may be understood as an exciton condensate analog of the Fulde-Ferrell-Larkin-Ovchinikov (FFLO) superconductor. We find that these direct exciton (D-ex) and ``FFLO exciton'' (FFLO-ex) orderings are often both present, but tend to compete, so that as one type of ordering increases the other shrinks.  With both present, the interlayer coherence should have spatial oscillations in real space.  In addition to this, intra-layer coherence between arcs on the same surface yields CDW order.

When CDW and exciton orders coexist, the system is in a supersolid state~\cite{Boninsegni_2012}.  Such order has been considered for bilayer systems in which Wigner crystals may form at low electron density~\cite{Zheng_1995, Narasimhan_1995,Joglekar_2006}, and tends to be associated with excitons localizing at sites in a two-dimensional crystal.  By contrast, in the coupled WSM surface system, the spatial ordering is determined by nesting vectors rather than by carrier density, so that there is no strong locking of the average inter-exciton separation with the CDW period.  Thus we expect the superfluid ordering to be more robust with respect to disorder than for the bilayer Wigner crystal system. A unique feature of the coupled FA system is the possibility of manipulating the relative strength of the spontaneous orderings by modifying the twist angle between surfaces, giving this system a level of tunability not present in more traditional materials. The presence of multiple continuously broken symmetries in this system implies that their superfluid modes will be coupled, so that counterflow superfluidity may become admixed with CDW sliding.  This could yield threshold behavior in counterflow supercurrent, above which narrow band noise is sustained.  Detection of such phenomenology associated with both exciton and CDW condensation would constitute direct evidence that the system hosts supersolid order.

The organization of the paper is as following. We introduce our model in Sec.~\ref{sec:Model}. In Sec.~\ref{sec:Greens_Function_and_Broken_Symmetry_States}, we describe the method for obtaining the interacting Green's functions. In Sec.~\ref{sec:Competing_phases}, we explore the competition between the phases of broken symmetry caused by capacitive coupling. In Sec.~\ref{sec:Goldstone_modes_and_counterflow_currents}, the Goldstone modes and counterflow current are discussed and  we conclude with a discussion and summary in Sec.~\ref{sec:Discussion_and_Summary}.

\begin{figure}[t]
	\centering
	\includegraphics[width=.95\linewidth]{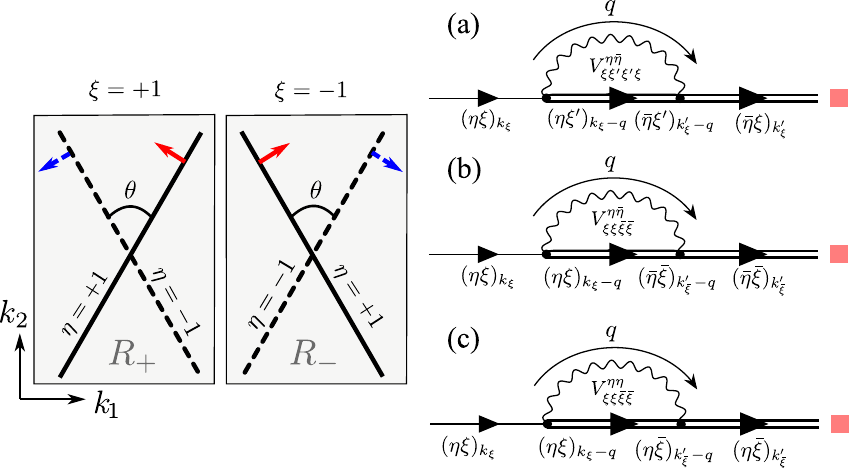}
	\caption{
		Left: Configurations of the FAs where solid and the dashed lines denote the states on the two surfaces ($\eta = \pm1$). $\xi=\pm1$ are two FAs on the same surface, at momentum region $R_{\xi}$. Directions of dispersions are marked by arrows. Pairs of FAs with $\xi\eta=\pm 1$ are nested with momentum $\vec{Q}$, which disperse in opposite directions.
		Right: Self-energy diagrams which describe spontaneously broken symmetries. $k_{\xi}, q$ represent four vectors, with momenta $\vec{k}_{\xi}\in R_{\xi}$. (a), (b) and (c) gives rise to self-energies for D-ex , FFLO-ex  and CDW  orders, respectively. \crule[red!50!white!100]{.2cm}{.2cm} stands for any of $(\eta,\xi)$.
		}
	\label{fig:selfen}
\end{figure}



\section{Model}
\label{sec:Model}
For concreteness, we consider a system of two WSMs with parallel surfaces labeled by an index $\eta=\pm 1$, a distance $t$ apart, with each surface hosting two FAs labeled by an index $\xi=\pm 1$, in general not parallel to one another [Fig.~\ref{fig:selfen}(a)].
Each FA joins the projections of two Weyl nodes onto the surface Brillouin-zone,
with wave functions that decay exponentially in the bulk of the WSM.   The
associated decay length diverges at the Weyl node projections, which we model by the inverse of a mass function $M_\xi(\vec{k})$~\cite{rkky}.
The single-particle energy associated with each arc has the form $\epsilon_{\xi}(\vec{k}) = \pm \hbar v_F k_{\perp}^{\xi}$, which disperses with the momentum component perpendicular to the ark, $k_{\perp}^{\xi}$.  Note the sign of this dispersion characterizes the helicity of the FA.  Further details of the model are provided in the Appendix \ref{app:construction_of_2_pair_of_tilted_fermi_arcs}.

We model interactions among the electrons by
\begin{align}
	H_{\rm int} = \sum_{\eta\eta'} \int_{\vec r,\vec r'} V^{\eta\eta'}(\vec r - \vec{r'}):\hat{\rho}^{\eta}(\vec{r})\hat{\rho}^{\eta'}(\vec{r'}):,\label{eq:hint}
\end{align}
where $\hat{\rho}^{\eta} = \sum_{\xi\xi'}\hat{\Psi}_{\xi}^{\eta\dagger}\hat{\Psi}_{\xi'}^{\eta}$ with $\hat{\Psi}^{\eta}_{\xi}$ being the field operator of the $(\eta,\xi)$ FA. The functions $V^{++}= V^{--}$ and $V^{+-}= V^{-+}$ are the intra- and inter-surface Coulomb interactions, respectively (Appendix \ref{sub:coulomb_matrix_elements}).
The decay depth of the single particle states entering our decomposition of the field operators $\hat{\Psi}^{\eta}_{\xi}$ impacts the matrix elements appearing when Eq.~(\ref{eq:hint}) is written in terms of the non-interacting FA states; beyond this, our model is two-dimensional.  We do not explicitly include bulk states in our analysis, although they can be approximately accounted for via screening in the interactions.

\section{Interacting Green's Function}
\label{sec:Greens_Function_and_Broken_Symmetry_States}
With these simplifications, components of the non-interacting finite-temperature Green's function of the WSM surfaces are given by $\mathcal{G}^0_{ij}(\vec{k},i\omega_n) = \delta_{ij}/(i\omega_n - \epsilon_{i}(\vec{k}))$, where $\omega_n = (2n+1)/k_B T$ are the fermionic Matsubara frequencies at temperature $T$ and the $i,j$ subscripts are composite indices for $\eta$ and $\xi$.
To describe the broken symmetry states, we include interactions through a self-energy matrix $\Sigma$, which introduces components in the Green's function even for $i\ne j$, as well as between different wave-vectors. These encode spontaneous ordering between different flavors of the fermions as well as possible translation symmetry breaking. We summarize the important diagrammatic terms that appear, within the non-crossing approximation~\cite{altland_book, Giuliani_book}, in Fig.~\ref{fig:selfen}(b).

In general, the full Green's function $\mathcal{G}$, non-interacting Green's function $\mathcal{G}^0$, and self-energy are related by the Dyson equation, to be solved self-consistently, $\mathcal{G} = \mathcal{G}^0 + \mathcal{G}^0 \Sigma \mathcal{G}$.
The diagrams illustrated in Fig.~\ref{fig:selfen}(b) represent the last term in this equation. Numerical solution of these requires integration over momentum, which we approximate as a discrete sum over
limited regions (denoted by $R_{\xi}$ in Fig.~\ref{fig:selfen}(a)) of the surface Brillouin zone in the vicinities of the FA's. In addition there is a Mastubara frequency sum, and the resulting self-energies are then independent of frequency because our model interaction is frequency independent. 
Further details of our numerical scheme are in the Appendix \ref{app:self_consistent_evaluation_of_self_energy}.

Denoting $\bar{\eta}=-\eta$ and $\bar{\xi}=-\xi$, and $k_{\xi}$ for the 3-momentum $(\vec{k}_{\xi}, \omega_n)$, we group the self-energies into three classes.
Coherence between FA's on different surfaces that do not spontaneously break translation symmetry have the form $\Sigma_{\xi\xi}^{\eta\bar{\eta}}(k_{\xi},k_{\xi}')$, and represent direct exciton order (D-ex) \cite{weylplsurf10}.  In addition, coherence between FA's on different surfaces which are separated in wavevector can also form, breaking both the gauge symmetry associated with individual layers and translational symmetry spontaneously, resulting in FFLO-ex order.  Such ordering is encoded in self-energy terms of the form $\Sigma_{\xi\bar{\xi}}^{\eta\bar{\eta}}(k_{\xi},k_{\bar{\xi}}')$. Finally, intra-layer interactions also give rise to self-energies of the form $\Sigma_{\xi\bar{\xi}}^{\eta\eta}(k_{\xi},k_{\bar{\xi}}')$, which indicate CDW order within a surface.  For simplicity, we neglect the diagonal terms in the self-energy $\Sigma_{\xi\xi}^{\eta\eta}$, which are expected to simply renormalize the non-interacting FA velocities.

\begin{figure}[t]
	\centering
	\includegraphics[width=.49\textwidth]{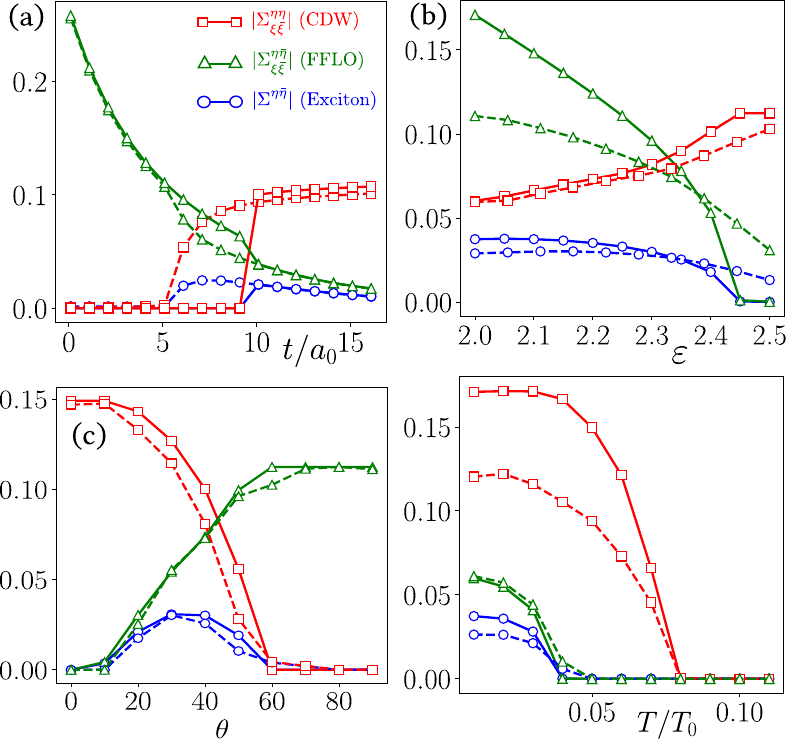}\\
	\caption{Behavior of maximum order parameter magnitudes with system parameters. Variation with: (a) surface separation ($t$); (b) WSM dielectric constant ($\varepsilon$); (c) FA tilt angle ($\theta$). Solid lines represent numerical results where the constraint Eq.~(\ref{eq:cons}) is enforced.  Results marked by dashed lines have these constraints relaxed. (d) Critical temperatures of the three order parameters, in  units of $T_0$ (see main text).
	$\theta=60^{\circ}$ is used in (a), (b) and (d). Other parameters: for (a), $\varepsilon = 2$; for (b), $t/a_0 = 5$; for (c), $t/a_0 = 5, \varepsilon = 2.5$; for (d), $t/a_0 = 5, \varepsilon = 2$.
	The separation of the Weyl nodes, as well as the nesting vector $|\vec{Q}|$, is taken to be $0.4$\AA$^{-1}$.}\label{fig:num}
\end{figure}

Before we discuss our numerical results, several comments are in order. Firstly, for the particular FA orientations shown in Fig.~\ref{fig:setup}, some FAs have parallel sections with nesting momentum $\vec{Q}$.  Particularly for the FFLO-ex order, one expects the dominant contribution of $\Sigma_{\xi\bar{\xi}}^{\eta\bar{\eta}}(\vec{k_{\xi}},\vec{k}_{\bar{\xi}}')$ to occur at
\begin{equation}
\label{eq:cons}
\vec{k}_{\xi}-\vec{k}_{\bar{\xi}}' = \pm\vec{Q}.
\end{equation}
Our numerical findings verify this for both FFLO-ex and CDW orders.
Numerical calculations can be greatly simplified by assuming these self-energies vanish except at this momentum difference.  We have compared this to results where the momentum difference is unconstrained in a few representative cases, and find rather good agreement~(see Fig.~\ref{fig:num} and Appendix \ref{sub:with_different_dispersion_direction}).
Secondly,
we characterize the strength of this Coulomb interaction by an effective
fine-structure constant, $\alpha = (c/\varepsilon v_F)/137$. Unless otherwise specified, we assume $\alpha = 5/\varepsilon$, which is consistent with $v_F\sim 10^5$m/s.  For the intra-surface interaction we adopt a dielectric constant $\varepsilon= 4$, while for the inter-surface interaction it is fixed at unity. We make the wave-vectors and lengths unitless in terms of a lattice-spacing distance ($a_0$) and we consider the unit of our energy-scale to be $T_0 = \hbar v_F/a_0$. With $a_0\sim 5$\AA, we have $T_0$ of the order of $10^3$K.


\color{black}

\section{Competing phases}
\label{sec:Competing_phases}

The numerical results we obtain indicate an intricate competition among the order parameters discussed above. Some typical results are illustrated in Fig.~\ref{fig:num}. At large $t$, for which interactions between surfaces are weak, intra-layer CDW order dominates, while for smaller separation and stronger inter-surface coupling, we find the FFLO-ex order dominates the CDW order. The competition between them is clearly visible in Figs.~\ref{fig:num}(a) and (b), in which we vary the separation between layers, and hence the relative intra- and inter-layer interactions.  Interestingly, at small separation only the FFLO-ex order is present, but with increasing separation a transition occurs in which D-ex and CDW orders set in, accompanied by a sharp drop in the FFLO-ex order.  This demonstrates the competition among the different types of order the system supports.  Note that for much of the parameter regime, the simplifying assumption expressed in Eq.~(\ref{eq:cons}) yields results largely consistent with calculations where this constraint is relaxed, except in the transition region, where it is necessary to relax the constraint to correctly capture its second order nature.

For fixed separation, it is notable that increasing the tilt angle between the arcs tends to enhance the FFLO-ex order at the expense of the D-ex and CDW orders, as shown in Fig.~\ref{fig:num}(c). This is clearly a consequence of the strong nesting between the arcs involved in the FFLO-ex ordering, which persists at all angles.  The relatively stronger stability of the this ordering is also apparent in the temperature dependence of the order parameters, illustrated in Fig.~\ref{fig:num}(d), indicating different critical temperatures for FFLO-ex and CDW orders. Interestingly we find that the critical temperature of the direct exciton order coincides with that of the intra-layer CDW order. At low temperatures, all three orders may coexist, and with rising temperature a transition may take place from such a \textit{multiply-ordered phase} to a phase with only FFLO-ex order. This intricate interplay of competition and cooperation among the different possible broken symmetries is one of the central results of this work.

Note that the FA dispersions shown in the Fig.~\ref{fig:selfen}(a) are oriented so that nested FAs of opposite surfaces disperse in opposite directions.  This supports the FFLO-ex order.  One may also consider situations in which they disperse in the {\it same} direction.  This could occur, for example, in WSM's with bulk magnetizations with opposite orientations.  This results in the loss of FFLO-ex order.  Introduction of curvature in the FA's also tends to suppress this order, although does not eliminate it (see Appendix~\ref{app:Tcwtheta}).

\section{Goldstone modes and counterflow currents}
\label{sec:Goldstone_modes_and_counterflow_currents}
Our model system involves four flavors of fermions (specified by $\eta$ and $\xi$), and each has an individually conserved charge that is encoded by a $U(1)$ symmetry.
The mean-field ground states we find spontaneously break at most three of these symmetries, so that all our phases respect global charge conservation.
Although in general the six self-energy terms ($\Sigma_{\xi\xi'}^{\eta\eta'}$, excluding $\xi=\xi'$ and $\eta=\eta'$) may attain non-zero values, their phases, $\theta_{\xi\xi'}^{\eta\eta'}$, are not independent. A close examination of the equations for the self-energies reveals the relations, $\theta^{+-}_{\xi\xi} + \theta^{++}_{\bar\xi\xi} = n_1 \pi,
~ \theta^{+-}_{\xi\xi} + \theta^{-+}_{\xi\bar\xi} = n_2 \pi,
~\theta^{\eta\bar\eta}_{-+} + \theta^{\eta\eta}_{+-}= n_3 \pi$,
where repeated indices are summed and the $n_i$'s may be $0$ or $\pm 1$ depending on the parameters.  Fluctuations of the phases that violate the above relations are massive, but variations which keep these relations intact increase the energy of the system only when they have spatial or temporal gradients.  Thus we expect our system to support three gapless Goldstone modes.  As detailed in the Appendix~\ref{app:Goldstone_modes_and_associated_supercurrents}, one may formally derive an effective action for phase fluctuations valid for small gradients in terms of three independent phases, $\theta_i$ ($i=0,2,3$), $S[\{\theta_i\}] \approx \sum_{q} \sum_{m,n} \Pi_{mn} (q)\theta_{n}(q)\theta_m(-q)$,
where $\Pi_{nm}(q)$ is the polarizability function. The normal modes of $S$ represent gapless modes, $\tilde{\theta}_i$, which are linear combinations of $\theta_i$. Static spatial gradients in these phases generally represent supercurrents,
$	(j_{\xi}^{\eta})_{l} = \sum_{i=1,2,3}\Gamma_{\eta \xi i}^{l}\nabla_l\tilde{\theta}_i$.  We present details of the form of $\Gamma$ for a simplified model in Appendix~\ref{app:Goldstone_modes_and_associated_supercurrents}; in general it depends on details of the system parameters and broken symmetries encoded in the $\Sigma$ matrix.  The entangling of different types of supercurrents, usually associated with interlayer counterflow currents \cite{Eisenstein_2014} or sliding CDW modes \cite{Gruner_1988}, is an important signature that in the generic case the ground state of this system is a {\it supersolid.} Remarkably,
for inversion-symmetric cases, we find a sum rule, $\sum_{\xi\eta} (j_{\xi}^{\eta})_{l} =0$,
indicating that the system does not support charged supercurrents.  While natural for particle-hole condensates, which support counterflow supercurrents, this is less obvious for CDW dynamics which support sliding modes.  We discuss the implications of this below.

\section{Discussion and Summary}
\label{sec:Discussion_and_Summary}
We have demonstrated that parallel surfaces of WSM's, with each hosting multiple FAs, in general support broken symmetries within and between the surfaces.
In particular, we show FFLO-exciton order may completely suppress direct-exciton and CDW orders, or may coexist with them.  In the latter case the system is a supersolid.  The entangling of orders in such a system is evidenced by its Goldstone modes, which in general have mixed counterflow - sliding CDW characters.  In real systems, sliding behavior of CDW's are not observed as a dissipationless current, because their broken translational symmetry necessary implies they will become pinned by disorder.  Nevertheless, they host unique transport signatures: threshold driving fields above which the a CDW may depin, and narrow-band noise with frequency proportional to the current above threshold \cite{Gruner_1988}.  An interesting signature of the supersolid character of this system would be the observation of these signatures in a counterflow experiment.

Several materials represent potential candidates for the physics described in this study. These include spinel compounds (such as VMg$_2$O$_4$)~\cite{spinel} and cobalt-based semimetals (such as Co$_3$Sn$_2$S$_2$). The former has two FAs on (110) surfaces, which are non-colinear and may serve as potential hosts for the physics we describe. For certain surface terminations, Co$_3$Sn$_2$S$_2$ has three FAs  which are oriented at 120$^{\rm o}$ angle with each other. For two such surfaces oriented at
$\sim 180^{\circ}$ one will have four FAs in approximately the configuration we consider; the other two may support their own FFLO-exciton condensation, but will essentially decouple from the other four FA's (see Appendix~\ref{app:3_fermi_arcs}).

For simplifications of the numerical analysis, we considered straight FAs for the non-interacting WSM surfaces. As argued in the Ref.~\onlinecite{excitonWSM}, in presence of a curvature in the FAs, 
there is an associated first order phase-transition with increasing curvature of the FAs. A full solution for the interacting Green's function in this case is numerically challenging; some results are presented in the Appendix. Moreover, in our idealization of these systems we have ignored the presence of bulk states which may be present at the Fermi energy, and can have finite support at the surfaces. Their impact on the broken symmetry states and associated supercurrents are interesting subjects for further study of these remarkably rich systems.

\color{black}

	

%


\begin{acknowledgments}
A.K  acknowledges support from the SERB (Govt. of India) via saction no. ECR/2018/001443, DAE (Govt. of India ) via sanction no. 58/20/15/2019-BRNS, as well as MHRD (Govt. of India) via sanction no. SPARC/2018-2019/P538/SL. H.A.F acknowledges support from the National Science Foundation via grant nos. ECCS-1936406 and DMR-1914451, as well as the support of the Research Corporation for Science Advancement through a Cottrell SEED Award, and the US-Israel Binational Science Foundation through award No. 2016130. We also  acknowledge the use of HPC facility at IIT Kanpur.
\end{acknowledgments}

\appendix

\onecolumngrid
\section{Construction of 2-pair of tilted Fermi arcs}%
\label{app:construction_of_2_pair_of_tilted_fermi_arcs}
Here we briefly discuss the way we construct multiple Fermi-arcs (FA), starting from one FA, using rotation and translation in momentum space. Let us consider a single FA as a straight line segment joining $(0,-k_0)$ and $(0,k_0)$ in the surface Brillouin-zone, with dispersion $\epsilon(\vec{k}) = \pm \hbar v_F k_x$, where $\pm$ denotes helicity. The states on the FA are supported primarily by the surface and decay exponentially in the bulk, where the inverse of the decay length is given by a mass function which we take to be $M(\vec{k}) = (k_0^2 - k_y^2) / 2 k_0$. We apply translation and rotation on this arc to generate FAs with other orientations on the momentum space. The translation and rotation matrices, which take the momentum $(k_x,k_y) \equiv (k_x,k_y,1)^T$, to a rotated and translated point $(k_x',k_y') \equiv (k_x',k_y',1)^T$, are given by
\begin{align}
	\label{eq:transformations}
	R(\theta) =
	\begin{pmatrix}
		\cos \theta & -\sin \theta & 0 \\
		\sin \theta & \cos \theta & 0 \\
		0 & 0 & 1 \\
	\end{pmatrix},~~
	T(u,v) =
	\begin{pmatrix}
		1 & 0 & -u \\
		0 & 1 & -v \\
		0 & 0 & 1 \\
	\end{pmatrix},
\end{align}
where the third direction is the perpendicular direction to the surface momenta introduced only for the mathematical operations and should not be considered after the operation is performed. Here $R(\theta)$ rotates the FA by an angle $\theta$ in the clockwise direction and $T(u,v)$ translates the origin of the FA from $(0,0)$ to $(u,v)$. With the composition of these two transformations one can generate the dispersions for multiple FAs.\\

\noindent
For the case of four FAs, which we focus on in the main text, one finds
\begin{align}
	\label{eq:rotated-dispersion}
	\epsilon_{\xi}^{\eta}(\vec k) &= -\xi \epsilon(R(\eta\theta /2)T(-\xi Q/2,0)\vec k),
						  & M_{\xi}^{\eta}(\vec k) &= M(R(\eta\theta /2)T(-\xi Q/2,0)\vec k).
\end{align}
In these expressions
$\vec{Q}$ is the separation of the FAs centers in momentum space (see Fig. 1 of the main text). This same method is used to generate the dispersions and mass functions for the $C_3$ symmetric
system discussed below in Appendix~ \ref{app:3_fermi_arcs}.

\section{Iterative solutions of the interacting Green's function}%
\label{app:interacting_hamiltonian}
In this section we provide more details of the construction of the self-energies and the equations we solve numerically for evaluating the interacting Green's function iteratively, focusing on the case of two Fermi-arcs on each of the WSM surfaces. A similar discussion for the hexagonal system with three Fermi-arcs on each of the surfaces is given in the Appendix~ \ref{app:3_fermi_arcs}. We take two FAs (indexed by $\xi=\pm1$) on each surface ($\eta=\pm1$) in the momentum region $R_{\xi}$ (see the main text). The wavefunctions of the FA states are
\begin{align}
\label{wavefunctions}
	\psi^{\eta}_{\xi , \vec k}(\vec r) = \frac{\exp(i \vec k \cdot \vec r)}{\sqrt{L_x L_y}} \sqrt{M_{\xi}^{\eta}(\vec k)} \exp(-\eta M_{\xi}^{\eta}(\vec k) z / 2) \Theta(\eta z) \Theta(M_{\xi}^{\eta}(\vec k)).
\end{align}
The density operator for the $\eta^{\rm th}$ surface is written as,
\begin{align}
	\label{eq:field-layer1}
	\hat{\rho}^{\eta}(\vec r) = \sum_{\xi\xi'}\hat{\Psi}_{\xi}^{\eta\dagger}(\vec r) \hat{\Psi}_{\xi'}^{\eta}(\vec r),
\end{align}
where, $\hat\Psi_{\xi}^\eta$ is the field operator for $(\eta\xi)^{\rm th}$ FA, which is constructed from the individual FA wave functions. In terms of these densities the Coulomb interaction is 
\begin{align}
	\label{eq:density-density-coulomb}
	H_{\rm int} = \sum_{\eta,\eta'}\int d{\vec r} \; d\vec{ r'} V^{\eta\eta'}(\vec r - \vec{r'}) :\hat{\rho}^{\eta}(\vec r) \hat{\rho}^{\eta'}(\vec{ r'}):.
\end{align}
Here, $V^{+-},\, V^{-+}$ are the inter-layer Coulomb interaction and $V^{++},\,V^{--}$ are
the intra-layer Coulomb interaction.  Substituting Eqs. \ref{wavefunctions} and \ref{eq:field-layer1} into Eq. \ref{eq:density-density-coulomb} leads to a two-dimensional Fourier transform of the Coulomb potential, which has the explicit form
$ V_{\vec q}^{\eta \eta'}(z - z') = {2 \pi \alpha e^{ - q |z - z'|}}/{ \varepsilon_{\eta \eta'} q} $, where $\vec{q} =(q_x,q_y)$. 
In units where the velocity of electrons $v_F$ is taken to be one, the effective fine structure constant characterizing the strength of the interaction is
$\alpha= \frac{c}{137 v_F}$, where $c$ is the speed of light in vacuum. The dielectric screening function $\varepsilon_{\eta\eta'}$ will in general be different between electrons in the same layer and in different layers. For our numerical modeling we take $\alpha=5$ and $\varepsilon_{++}=\varepsilon_{--}=\varepsilon$, which we vary as a system parameter, whereas we keep $\varepsilon_{+-}=\varepsilon_{-+}=1$.  Thus in exploring the effect of the relative intra- and inter-layer interaction strengths, we keep
the dielectric constant of the spacer layer fixed at unity and vary that of the WSM slabs via $\varepsilon$.

\subsection{Coulomb Matrix elements}%
\label{sub:coulomb_matrix_elements}
The Coulomb matrix elements are given by
\begin{align}
	V_{\xi_1\xi_2\xi_2\xi_3}^{\eta \eta'}(\vec k, \vec k', \vec q) &= \int dzdz'
	V_{\vec q}^{\eta\eta'}(z-z')
	\rho_{\xi_1\xi_2}^{\eta}(z + (\delta_{\eta,-1} - \delta_{\eta,+1})t / 2,\vec k - \vec q, \vec k) \rho_{\xi_3\xi_4}^{\eta'}(z' + (\delta_{\eta',-1} - \delta_{\eta,+1})t / 2,\vec k + \vec q, \vec k), \nonumber
\end{align}
where
\begin{align}
		\rho_{\xi_1\xi_2}^{\eta}(z,\vec k', \vec k)
	&=
	\sqrt{M^{\eta}_{\xi_1}(\vec k)
		M^{\eta}_{\xi_2}(\vec k')}
	\exp( \frac{-\eta z}{2}(M^{\eta}_{\xi_1}(\vec k)+M^{\eta}_{\xi_2}(\vec k')))
	\Theta\left(M^{\eta}_{\xi_1}(\vec k)\right)
	\Theta\left(M^{\eta}_{\xi_2}(\vec k')\right)
	\Theta\left(\eta z\right),
\end{align}
with $t$ the thickness of the medium between the WSM's. One then finds
\begin{align}
&	V_{\xi_1\xi_2\xi_3\xi_4}^{\eta\eta'}(\vec k, \vec k', \vec q) =
	\frac{8 \alpha \pi}{ \varepsilon_{\eta\eta'} q} e^{-(1-\delta_{\eta,\eta'})q t}
	\sqrt{M^{\eta}_{\xi_1}(\vec k) M^{\eta}_{\xi_2}(\vec k - \vec q)
		M^{\eta'}_{\xi_3}(\vec k') M^{\eta'}_{\xi_4}(\vec k' + \vec q)
	}\times
	\nonumber\\
	&~~~\times \frac{\Theta\left(M^{\eta}_{\xi_2}(\vec k - \vec q)\right)
		\Theta\left(M^{\eta'}_{\xi_4}(\vec k' + \vec q)\right)
		\Theta\left(M^{\eta'}_{\xi_3}(\vec k')\right)
		\Theta\left(M^{\eta}_{\xi_1}(\vec k)\right) }{(M^{\eta}_{\xi_2}(\vec k - \vec q) + M^{\eta}_{\xi_1}(\vec k)+ 2q)(M^{\eta'}_{\xi_4}(\vec k' + \vec q) + M^{\eta'}_{\xi_3}(\vec k') + 2q)} \left[1 + \frac{2 q \delta_{\eta\eta'}}{M^{\eta}_{\xi_2}(\vec k - \vec q) + M^{\eta}_{\xi_1}(\vec k) +M^{\eta'}_{\xi_4}(\vec k' - \vec q) + M^{\eta'}_{\xi_3}(\vec k') } \right].	\label{eq:inter-layer}
\end{align}
Note that
the inter-layer Coulomb matrix elements are exponentially suppressed with increasing momentum due to the separation between the two layers, while the intra-layer Coulomb matrix elements contain a dielectric constant ($\varepsilon$) that approximately captures the screening due to electrons in the bulk.

\begin{figure}[t]
	\centering
	\includegraphics{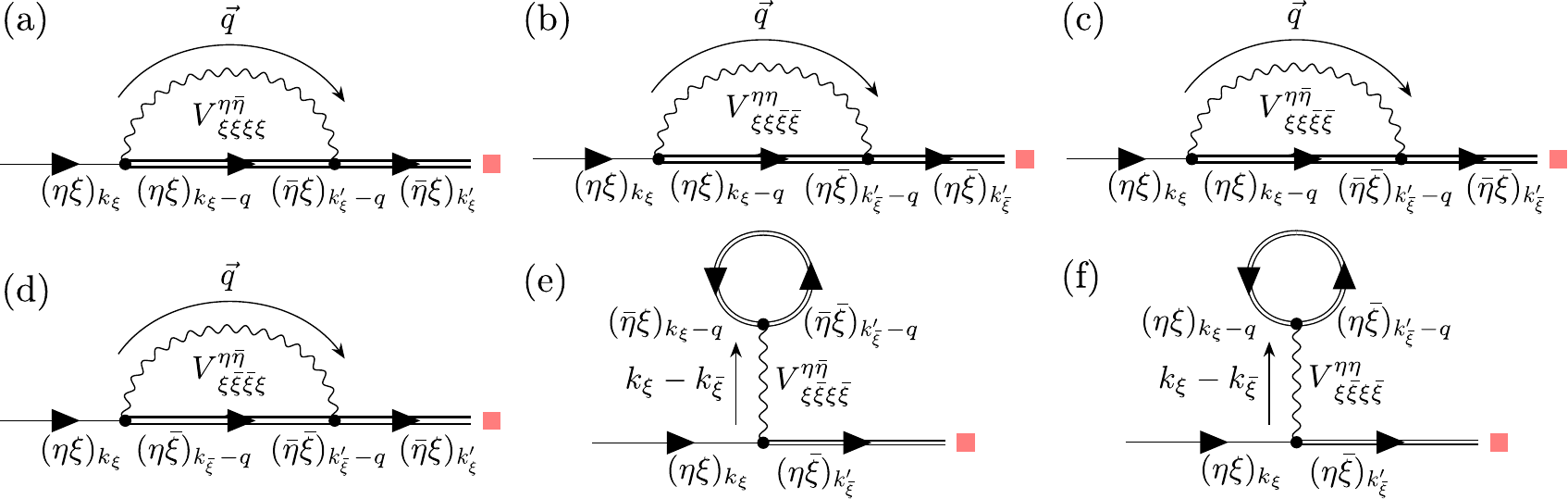}
	\caption{
		All possible self-energy insertion consistent with momentum
		conservation.  The diagrams (a), (c), (d), (e) are generated by
		interlayer interaction, and in (a), (d) the self-energy is excitonic,
		in (c) the self-energy terms is FFLO type, and in (f) the self-energy
		contributes to CDW order. Intra-layer interaction leads to CDW type
		order only, as (b) and (f).  \crule[red!50!white!100]{.2cm}{.2cm}
		stands for any $(\eta\xi)$, consistent with momentum conservation.
	}%
	\label{fig:SE}
\end{figure}

\subsection{Self consistent evaluation of self energy}%
\label{app:self_consistent_evaluation_of_self_energy}

With these Coulomb matrix elements, we can consider self-energy corrections to the electron Green's function of this system. Because we are interested in broken symmetry states, it is necessary to adopt a self-consistent approximation that captures contributions at all orders in perturbation theory in the interaction.  The self-energy form we adopt is a sum of all non-crossing diagrams up to infinite order (``non-crossing approximation''~\cite{altland_book}). In cases where translation symmetry can be broken (FFLO and CDW orders) we also include tadpole diagrams, although the large momentum exchanges involved (of the order $\sim Q$, which is the separation of the FAs in momentum space) suppresses their contributions, so that they have little effect on the final result. We enumerate all the self-energy contributions we retain
in Fig.~\ref{fig:SE}. These can be organized into three classes: (i) Direct exciton self-energy diagrams
(Figs.~\ref{fig:SE}(a), (d)); (ii) CDW
self-energy diagrams (Figs.~\ref{fig:SE}(b), (e), (f)); and
(iii) FFLO exciton self-energy diagrams.  Note that the momentum carried by the interactions in Figs. \ref{fig:SE}(e) and (f) is essentially the nesting vector $Q$.

The internal Green's function lines in the self-energies are given by the Dyson equation, which is usefully written in the form ${\mathcal G} = (({\mathcal G}^0)^{-1} - {\Sigma})^{-1}$.   Because the system consists of multiple Fermi arcs,  the interacting Green's function and the self-energies have a matrix structure.  We take the latter of these to be of the form
\begin{align}
	\Sigma(\vec k, \vec{k'}) &=
	\begin{pmatrix}
		0 & ~\Sigma^{+-}_{--}~ &  ~\Sigma^{++}_{-+}~ & ~\Sigma^{+-}_{-+}~ \\
		~\Sigma^{-+}_{--}~ & 0 & ~\Sigma^{-+}_{-+}~ &  ~\Sigma^{-+}_{-+}~\\
		~\Sigma^{++}_{+-}~ & ~\Sigma^{+-}_{+-}~ & 0 & ~\Sigma^{+-}_{++}~ \\
		~\Sigma^{-+}_{+-}~ & ~\Sigma^{--}_{+-}~ & ~\Sigma^{-+}_{-+}~ & 0
	\end{pmatrix}_{(\vec k, \vec{k'})},\nonumber
\end{align}
where the non-vanishing self-energy matrix elements are shown in Figs.~\ref{fig:SE}, and we have neglected the diagonal contributions $\Sigma_{\xi\xi}^{\eta\eta}$, which typically can be incorporated into velocity renormalizations of the FA's. The non-interacting Green's functions are taken to have the form $\mathcal{G}^0_{ij}(\vec{k},i\omega_n) = \delta_{ij}/(i\omega_n - \epsilon_{i}(\vec{k}))$,  where $i,j$ index the FA's, and $\omega_n$ are the Matsubara frequencies.

As illustrated, the self-energy diagrams in 
Fig.~\ref{fig:SE} assume the single-particle states are characterized by single wavevectors, 
and generally one may find self-consistent solutions for the Green's functions in which a state at $\vec{k}$ admixes with a continuous set of momenta $\vec{k}'$.  
For this general situation, the self energies take the explicit forms
\begin{align}
	\label{eq:eqnOne}
	\Sigma^{\eta\bar\eta}_{\xi\xi}(\vec k_{\xi}, \vec{k_{\xi}'}) &= -\frac{1}{\beta} \sum_{\vec q}\sum_{n} \Big[
	V^{\eta\bar\eta}_{\xi\xi\xi\xi}(\vec k_{\xi}, \vec k_{\xi}', \vec q)
	\mathcal G^{\eta\bar\eta}_{\xi\xi}(\vec k_{\xi} -\vec q,\vec k_{\xi}' -\vec q, \omega_n - \omega_m) \\
												 &~~~+ V^{\eta\bar\eta}_{\xi\bar\xi\bar\xi\xi}(\vec k_{\xi}, \vec k_{\xi}', \vec q) \mathcal G_{\bar\xi\bar\xi}^{\eta\bar\eta}(\vec k_{\bar\xi} -\vec q,\vec k_{\bar\xi}' -\vec q, \omega_n - \omega_m) \Big]e^{i( \omega_n - \omega_m )0^{+}}\nonumber ,\\
	\Sigma^{\eta\bar\eta}_{\xi\bar\xi}(\vec k_{\xi}, \vec k_{\bar\xi}') &= -\frac{1}{\beta} \sum_{\vec q}\sum_{n} V^{\eta\bar\eta}_{\xi\xi\bar\xi\bar\xi}(\vec k_{\xi} , \vec k_{\bar\xi}', \vec q) \mathcal G_{\xi\bar\xi}(\vec k_{\xi} -\vec q,\vec k_{\bar\xi}' -\vec q, \omega_n - \omega_m) e^{i( \omega_n - \omega_m )0^{+}} ,\\
	\Sigma^{\eta\eta}_{\xi\bar\xi}(\vec k_{\xi}, \vec k_{\bar\xi}') &= \frac{1}{\beta} \sum_{\vec q}\sum_{n} \Big[- V^{\eta\eta}_{\xi\xi\bar\xi\bar\xi}(\vec k_{\xi}, \vec k_{\bar \xi}', \vec q) G_{\xi\bar\xi}(\vec k_{\xi} -\vec q,\vec k_{\bar \xi}' -\vec q, \omega_n - \omega_m) e^{i( \omega_n - \omega_m )0^{+}}
																 \\&~~~+ V^{\bar\eta\bar\eta}_{\xi\bar\xi\xi\bar\xi}(\vec k_{\xi}, \vec k_{\bar\xi}' - \vec q, \vec k_{\xi} - \vec k_{\bar\xi}') \mathcal G^{\bar\eta\bar\eta}_{\bar\xi\bar\xi}(\vec k_{\bar\xi}' - \vec q, \vec k_{\xi} - \vec q)e^{i \omega_n 0^+} \nonumber
																 \\&~~~+ V^{\eta\eta}_{\xi\bar\xi\xi\bar\xi}(\vec k_{\xi}, \vec k_{\bar\xi'} - \vec q, \vec k_{\xi} - \vec k_{\bar \xi}') G_{\bar\xi\xi}(\vec k_{\bar\xi}' - \vec q, \vec k_{\xi} - \vec q)e^{i \omega_n 0^+}
		\Big].	\label{eq:eqntwo}
\end{align}
In the most general case, we allow all possible pairs of momenta ($\vec k$,$\vec k'$) within fixed regions of the FA's. Numerically the Green's function then becomes a $4N_k \times 4N_k$ dimensional matrix, where $N_k$ is the number of momentum grid points we retain in the numerical calculations. As discussed in the main text, due to the strong nesting with  wavevector $\vec Q$ between some of the Fermi arcs, we find that dominant contributions to the CDW and FFLO-exciton self-energies correspond to $\vec k - \vec k' = \vec Q$, while that of the direct-exciton self-energies correspond to $\vec{k}=\vec{k'}$. This observation allows us to work with a $4\times 4$ self-energy matrix for each momentum $\vec{k}$, greatly reducing the numerical complexity of the problem.  In either case we perform the Matsubara summation analytically by contour integration (the poles of the interacting Green's function are found numerically), followed by a summation over the internal momenta, with an energy-cutoff of $|\epsilon_{\rm cut} | = 0.5$ and a momentum grid of 1604 number of points have been used.  The equations are solved iteratively until convergence up to a tolerance of order $10^{-6}$ is reached.  As we are not considering dynamical screening of the Coulomb interaction, after summation in \eqref{eq:eqnOne}--\eqref{eq:eqntwo}, the resultant self-energies are frequency-independent, and we assuming that the converged solutions do not depend strongly on the energy or the momentum cutoff.

\begin{figure}[t]
	\centering
	\includegraphics[width=\textwidth]{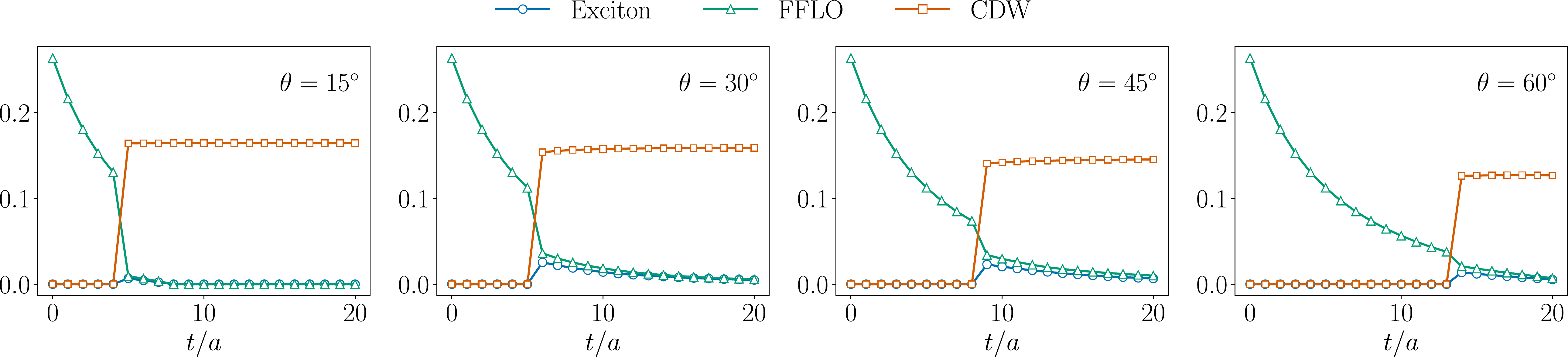}
	\caption{
		The maximum amplitude of self-energy terms are shown as a function of
		distance between the two slabs for four different tilt angles between
		the arcs.
		For all of the figures, the separation of the Weyl nodes,
		$2k_0$ is taken to be $0.4$\AA$^{-1}$, the
		dielectric constant of the WSM slabs are $\varepsilon=2.35$, here length of the nesting vector is taken to be
		$\left|{\vec Q}\right| = 4k_0$. These results are obtained 4$\times$4 matrix valued Green's function.
	}
	\label{fig:t_plot-pdf}
\end{figure}

\begin{figure}[t]
	\centering
	\includegraphics[width=\textwidth]{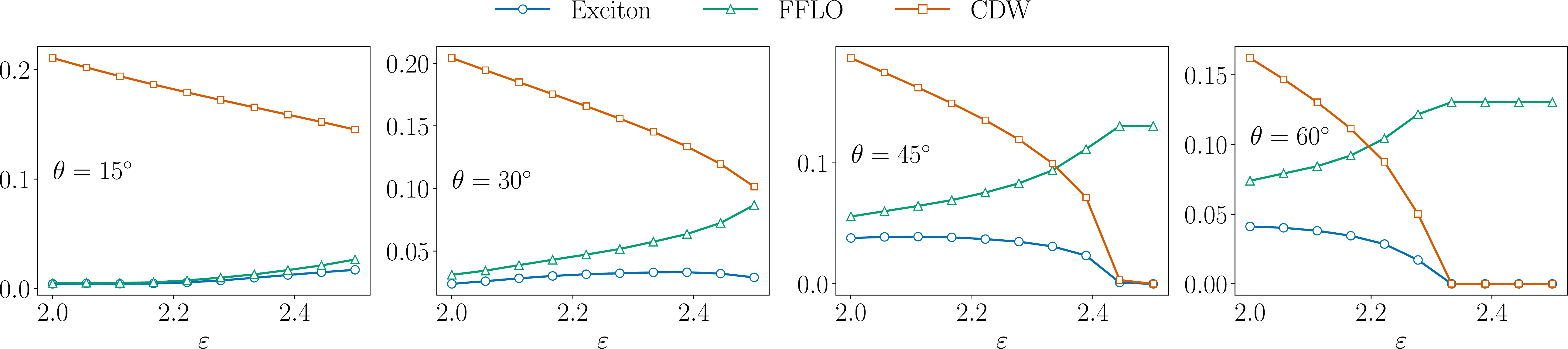}
	\caption{
		The maximum absolute values of self-energy terms is shown as a function of the dielectric constant of the WSM slabs.  For all of the figures, the separation of the Weyl nodes, $2k_0$ is taken to be $0.4$\AA$^{-1}$, the vertical separation of the WSM slabs are taken as $t / a=4$, the length of the nesting vector is taken as, $\left|{\vec Q}\right| = 4k_0$.
	}
	\label{fig:epsilon_plot}
\end{figure}

\begin{figure}[t]
	\centering
	\includegraphics[width=\linewidth]{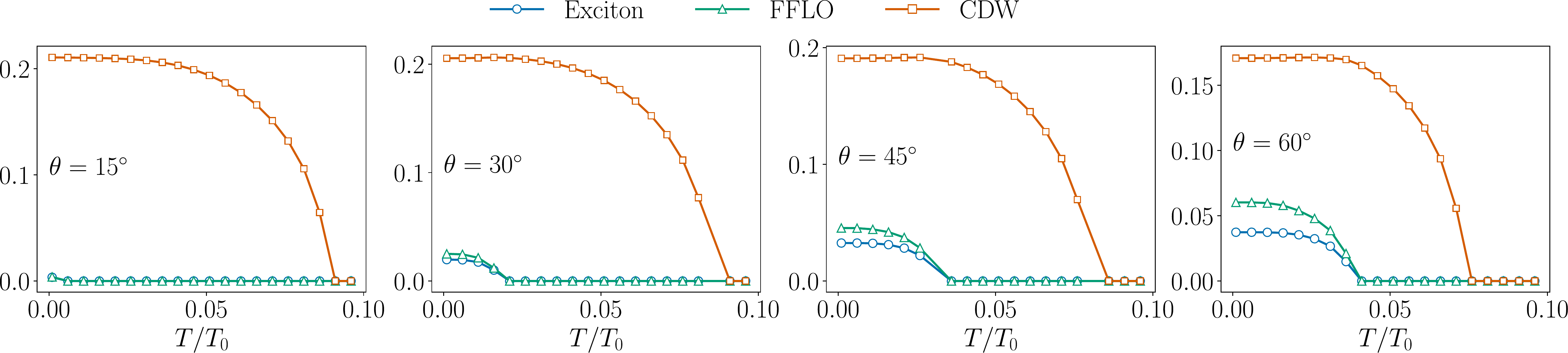}
	\caption{
		The maximum absolute value of self-energy terms with respect to
		temperature at different tilt angles $\theta$ is shown. With increasing
		temperature the curves follow a second order transition. A key
		observation is that critical temperature for CDW an exciton, FFLO order
		is different. For all of the figures, the separation of the Weyl nodes,
		$2k_0$ is taken to be $0.4$\AA$^{-1}$, the vertical separation of the
		WSM slabs are taken as $t / a=5$, temperature $T_0\approx10^3$K, the
		dielectric constant of the WSM slabs are $\varepsilon=2$, here
		$\left|{\vec Q}\right| = 4k_0$. For these set of parameters CDW order
		dominates, and with increasing tilt angle, the critical temperature of
		exciton and FFLO increases and that of CDW decreases. These results are obtained 4$\times$4 matrix valued Green's function.
	} %
	\label{fig:4-Tc}
\end{figure}

\section{Variation of self-energy with tilt angle and other system parameters}%
\label{app:Tcwtheta}
In this Section we present some further results
of how the self-energy evolves under changes of different parameters. We use the simplified simplified self-energy model described above, in which the Green's function is diagonal in momentum, and has a 4$\times$4 matrix for each value of $\vec{k}$.

\subsubsection*{Dependence on the separation between the WSM slabs $(t)$}
From Eq. \eqref{eq:inter-layer} one sees 
that the matrix elements of the inter-surface Coulomb interaction decays exponentially with $t$
(\(\sim e^{-q t}\)), in contrast to the intra-surface Coulomb interaction, which is independent of $t$. This allows for tuning of the relative inter- and intra-surface interactions, and their competition is
illustrated in Fig.~\ref{fig:t_plot-pdf}. As the separation between
the WSM slabs is increased, one observes that
state supports only FFLO-exciton order, and above some critical separation,  
CDW order sets in, the FFLO-exciton order is greatly suppressed, and direct-exciton order appears as well, and closely tracks the FFLO-exciton order.  Note that within our simplified single $\vec{Q}$ approximation the transition is first order, but a full calculation will likely convert this to a second order transition.   The relative strengths of the two types of orders are expected to be qualitatively the same as shown in Fig. \ref{fig:t_plot-pdf}. With larger tilt angle between the Fermi arcs, this critical thickness also increases.

\subsubsection*{Dependence on the dielectric constant \(\varepsilon\)}
From Eq.~\eqref{eq:inter-layer} we see that the intra-layer interaction
scales as \(\sim 1 / \varepsilon\), so with increasing the dielectric
constant of the WSM (or, equivalently, decreasing the dielectric constant of the spacer layer) the intra-layer interaction becomes smaller relative to the inter-layer interaction.  At large enough $\varepsilon$, the FFLO-exciton order becomes dominant over the other types of order, and for large enough values it is the only broken symmetry in the ground state. This behavior is shown in Fig.~\ref{fig:epsilon_plot}.

\begin{figure}[b]
	\centering
	\includegraphics[width=0.8\linewidth]{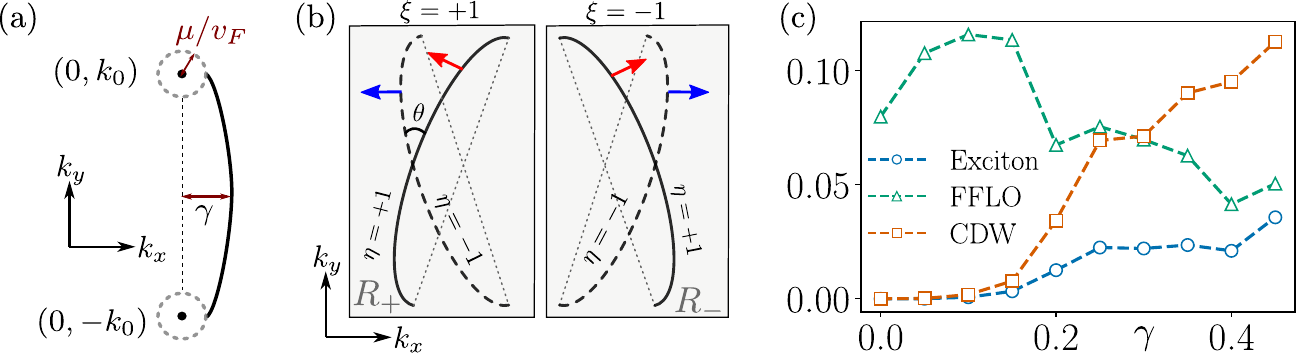}
	\caption{
		The curved-FA is shown in (a), with two Weyl nodes at $(0,\pm k_0)$, with curvature $\gamma$. The curved FAs always end on circles with radius $\mu / v_F$ for chemical potential $\mu$.  These circles represent a Fermi surface for bulk states. The four FA configuration is shown in (b), where the solid and dashed lines denote the states on the two surfaces ($\eta = \pm1$).  The directions of the energy dispersions are indicated by arrows. The dotted lines corresponding to straight FAs are drawn for comparison. The measure of curvature is denoted by $\gamma$ in \eqref{eq:curvFA-disp}.  (c) Dependence of self-energies on the curvature. As the curvature increases, due to the lack of nesting the FFLO-exciton self-energy decreases. For these numerical results the separation of the Weyl nodes $2k_0$ is taken to be $0.4$\AA$^{-1}$, the vertical separation of the WSM slabs is taken as $t / a=5$, the dielectric constant of the WSM slabs is $\varepsilon=2.5$, and the chemical potential is taken to be $\mu = 0.05$.
	}%
	\label{fig:curved_FA}
\end{figure}

\subsubsection*{Dependence on the temperature \( (T / T_0)\)}
The previous results suggest that direct-exciton and CDW orders are coupled in such a way that they grow or shrink together as system parameters vary, whereas the FFLO-exciton order evolves more independently.  However the interdependence of the orders is more complicated than this.  We illustrate this by considering the temperature dependence of the self-energies for a particular parameter regime, as shown in Fig.~\ref{fig:4-Tc}.
Interestingly, for this parameter range (in particular, the relatively small value of $\varepsilon=2$), the CDW order dominates at low temperature, and both types of exciton order are relatively suppressed.  With increasing temperature the exciton orders vanish together at a common transition temperature, while the CDW order persists up to a higher transition temperature.
Interestingly, with increasing tilt angle, the critical temperature of the inter-surface orders  increases, while that of the CDW order decreases. $T_0$ is defined in the main text.

\subsubsection*{Dependence on the curvature of the Fermi Arc}
In the main text, the FA's we considered were straight line segments, whereas in real materials FAs are often curved. 
To examine this we model the dispersions of curved Fermi arcs as
\begin{align}
	\epsilon(\vec k) = \pm v_F k_x \left(1- \gamma \sqrt{1-\left(\frac{k_y}{k_0}\right)^2}\right).
	\label{eq:curvFA-disp}
\end{align}
As shown in shown in Fig.~\ref{fig:curved_FA}(a), at finite chemical potential
$\mu$, the parameter $\gamma$ measures the curvature of the Fermi arcs. The Fermi arcs also
touch Fermi circles of radius $\mu / v_F$ around the Weyl points $(0,\pm k_0)$,
which denote bulk states. Notice that when $k_x$ passes through zero and the dispersion changes sign, the sign of the FA curvature also changes, a remarkable resemblance to realistic curved FA's. We generate four tilted FAs in the same way as described in Appendix~~\ref{app:construction_of_2_pair_of_tilted_fermi_arcs}, resulting in the geometry illustrated in Fig.~\ref{fig:curved_FA}(b), and our numerical scheme for computing self-energies is essentially the same as for the case of straight FAs.  As the curvature increases the nesting between the Fermi arcs decreases, so that the 4$\times$4 Green's function approach becomes a poor approximation. We thus find self-consistent solutions for the self-energy terms without any constraints on the momentum.  

The full solution for the interacting Green's function in the presence of this curvature in the FAs is numerically challenging, so we consider results for only a single set of representative parameters in Fig.~\ref{fig:curved_FA}(c). As might be expected the FFLO-exciton order decreases with increasing curvature since perfect nesting is spoiled. Interestingly, this suppression with curvature is accompanied by an increase in CDW and direct-exciton orders, which we attribute to a combination of improved nesting of the coupled arcs in these latter orders, and the suppression of the competing FFLO-exciton order.  It is also possible that curvature may lead to onset of the spontaneous orders via first order transitions \cite{excitonWSM} in some regions of the parameter space, however we do not observe this behavior in Fig.~\ref{fig:curved_FA}(c).\\

\subsection{WSM magnetizations in the opposite directions}%
\label{sub:with_different_dispersion_direction}
\begin{figure}[ht]
	\centering
	\includegraphics[width=\linewidth]{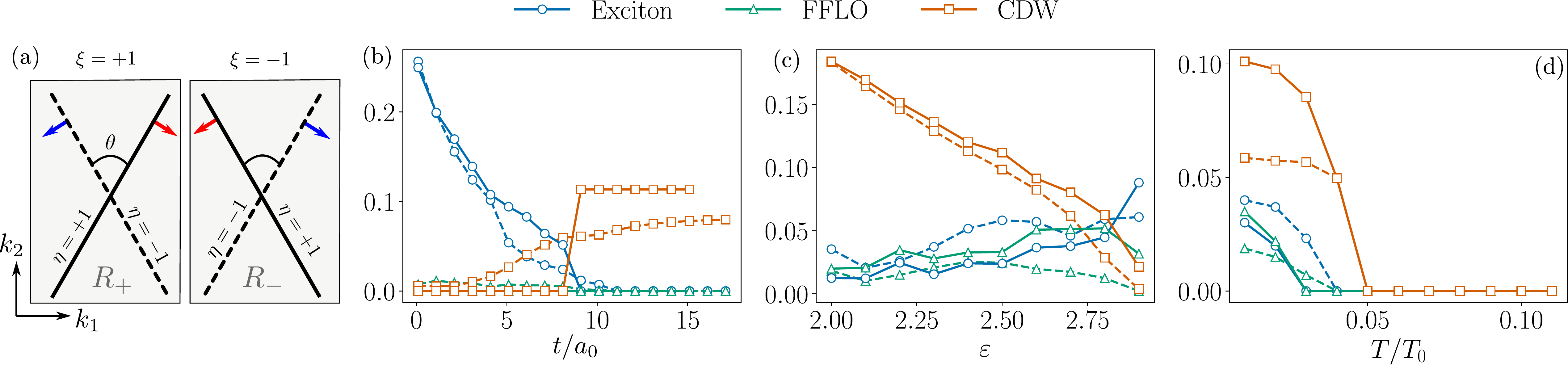}
	\caption{In (a) the case of 2-pair of Fermi arcs is shown with opposite magnetization. Unlike in the main text there is no nesting vector between arcs. The maximum magnitudes of the order-parameters as a function of system parameters such as (b) the separation ($t$) between WSMs and (c) the dielectric constant ($\varepsilon$) of the WSMs. The solid lines represent the numerical results where the constraint $\vec k_{-1} - \vec k_{+1} = \vec Q$ is used, in comparison to the results marked by dashed lines where such constraints have not been used. (d) Shows critical temperatures for the three order parameters which indicates the same critical temperature of the FFLO and excitonic orders. $\theta=60^{\circ}$ is used in (b), (c) and (d). Other parameters used are, for (b): $\varepsilon = 2.5$; for (c): $t/a_0 = 5$; for (d): $t/a_0 = 5, \varepsilon = 2.5$. For numerical integrations, the momentum-cut for individual FAs are taken using a energy cut-off of $|\epsilon_{\rm cut} | = 0.5$ and a momentum grid of 1604 number of points have been used. The separation of the Weyl nodes, $2k_0$, is taken to be of $0.4$\AA$^{-1}$, and $|\vec{Q}| = 2k_0$.}%
	\label{fig:SE-other-config}
\end{figure}
The coupled surfaces discussed in most of this work can be imagined as involving two WSM's with fixed bulk magnetizations.  In principle the magnetization of one of the two bulk systems may be reversed, inverting the FA dispersions on one of the two surfaces.  In this section we briefly examine the effect of such a reversal.
To do this we adopt FA dispersions of the form $\epsilon_{\xi}^{\eta}(\vec k) = \xi\eta \epsilon(R(\eta\theta /2)T(-\xi Q/2,0)\vec k).$
The dependence of the resulting self-energies on $t, \varepsilon$, and $T$, are illustrated in Fig.~\ref{fig:SE-other-config}.  One significant difference from the previous results is that, due to a lack of strong nesting in the FFLO-exciton channel, this type of ordering never dominates the various broken symmetries.  The dependence on various parameters are as follows.\\

\emph{Separation between the WSM slabs} ($t$)--
With increasing $t$ the direct-exciton self-energy magnitude, which dominates at small $t$, decays due to the exponential decay factor in the interlayer interaction as given in Eq.~\eqref{eq:inter-layer}. Beyond a critical thickness the CDW order starts to grow, and saturates as the direct-exciton self-energy magnitude becomes zero, as seen in Fig.~\ref{fig:SE-other-config}(b).  Thus we see in this situation that the CDW and direct-exciton tend to compete, in contrast to the situation where FFLO-exciton order dominates, for which these orders can grow together with changing parameters.\\

\emph{Dependence on dielectric constant} (\(\varepsilon\))--
With increasing the dielectric constant with the WSM slabs, the intra-layer interaction decreases and CDW order decreases as a result. Beyond a critical $\varepsilon$ the direct-exciton order starts to dominate as is shown in Fig.~\ref{fig:SE-other-config}(c).\\

\emph{Dependence on temperature} (\(T / T_0\))--
As a function of temperature we find a continuous transition to a disordered state, as the case for the other relative orientations of the bulk magnetizations. For $\theta = 60^\circ, t / a = 5, \varepsilon = 2.5$, we find that CDW order has the maximal critical temperature, as illustrated in Fig.~\ref{fig:SE-other-config}(d). In this case direct- and FFLO-exciton orders have same the critical temperature. $T_0$, the temperature scale of the system, is defined in the main text.
\color{black}



\section{Goldstone modes and associated supercurrents}
\label{app:Goldstone_modes_and_associated_supercurrents}
As mentioned in the main text, the bare Hamiltonian has four $\mathrm U(1)$ symmetries, associated with conservation of charge for each individual FA.  The mean-field ground state spontaneously breaks three $\mathrm U(1)$ symmetries, while the total charge of the system remains conserved.  Thus we expect the system to support three gapless Goldstone modes.
Even when the six self-energies ($\Sigma_{ij}$) attain non-vanishing mean-field values, their phases are not independent and we find relationships among them, given by


\begin{align}
	\sum_{\xi} \left[ \theta^{+-}_{\xi\xi} + \theta^{++}_{\bar\xi\xi}\right] = n_1 \pi,
	&& \sum_{\xi} \left[ \theta^{+-}_{\xi\xi} + \theta^{-+}_{\xi\bar\xi}\right] = n_2 \pi,
	&& \sum_{\eta} \left[ \theta^{\eta\bar\eta}_{-+} + \theta^{\eta\eta}_{+-}\right] = n_2 \pi
	\label{eq:phase1}
\end{align}
where we take $n_i =0$ in the following. The Green's function is written as
\def\thab{\theta^{+-}_{--}}
\def\thcd{\theta^{+-}_{++}}
\def\thad{\theta^{+-}_{-+}}
\def\thbc{\theta^{-+}_{-+}}
\def\thac{\theta^{++}_{-+}}
\def\thbd{\theta^{--}_{-+}}
\begin{align}
	\mathcal{G}^{-} =
	\begin{pmatrix}
		i \partial_t - \phi^{+}_{-} - \epsilon^{+}_{-}(\vec k - \vec A^{+}_{-}) & d_1 e^{i \thab} & d_2 e^{i \thac}& d_3 e^{i \thad} \\
		d_1 e^{-i \thab}& i \partial_t - \phi^{-}_{-} - \epsilon^{-}_{-}(\vec k - \vec A^{-}_{-}) & d_3 e^{i \thbc} & d_2 e^{i \thbd} \\
		d_2 e^{-i \thac}& d_3 e^{-i \thbc}& i \partial_t - \phi^{+}_{+} - \epsilon^{+}_{+}(\vec k - \vec A^{+}_{+}) & d_1 e^{i \thcd}\\
		d_3 e^{-i \thad}& d_2 e^{-i \thbd}& d_1 e^{-i \thcd} & i \partial_t - \phi^{-}_{+} - \epsilon^{-}_{+}(\vec k - \vec A^{-}_{+})
	\end{pmatrix},
	\label{eq:green}
\end{align}
where $d_1,d_2$, and $d_3$ are the amplitudes of the direct-exciton, CDW and FFLO-exciton self-energies, respectively.  We have introduced scalar a ($\phi_{\xi}^{\eta}$) and a vector ($\vec{A}_{\xi}^{\eta}$) potential for each of the FAs. Neglecting any variation of the self-energy amplitudes~\cite{altland_book}, and noting that any fluctuations of the phases that alter the above relations, Eq.~(\ref{eq:phase1}), will be massive, we consider only fluctuations of the phases which respect them. Thus there will be three gapless (Goldstone) modes. To implement this we retain three independent phases \( \thad, \thbd, \thcd\), and express all other phases in Eq. \ref{eq:green} in terms of them using Eqs.~\eqref{eq:phase1}. Introducing three independent phases ($\theta_1,\theta_2,\theta_3$),
\begin{align}
\thad = \theta_1+ \theta_3,~~ 	\thbd= \theta_1 - \theta_2,~~ \thcd = -\theta_2 + \theta_3,
\end{align}
we perform a unitary transformation $U\mathcal{G}^-U^{\dag}$, with
\begin{align}
	&U \equiv \mathrm{diag}(e^{-i /2 (\theta_1 + \theta_2 + \theta_3)}, e^{-i /2 (\theta_1 - \theta_2 -\theta_3)}, e^{i /2(\theta_1 + \theta_2 - \theta_3)}, e^{i / 2 (\theta_1 - \theta_2 + \theta_3)}).
\end{align}
This renders all off-diagonal entries of $G^-$ real, and introduces spatial and temporal fluctuations of the phases. One finds explicitly
\begin{align}
	U\mathcal G^{-}U^{\dag}&=
	\begin{pmatrix}
		i \partial_t - \tilde \phi^{+}_{-} - \epsilon^{+}_{-}(\vec k - \tilde {\vec A}^{+}_{-}) & d_1  & d_2 & d_3  \\
		d_1 & i \partial_t - \tilde \phi^{-}_{-} - \epsilon^{-}_{-}(\vec k - \tilde {\vec A}^{-}_{-}) & d_3  & d_2  \\
		d_2 & d_3 & i \partial_t - \tilde \phi^{+}_{+} - \epsilon^{+}_{+}(\vec k - \tilde {\vec A}^{+}_{+}) & d_1 \\
		d_3 & d_2 & d_1  & i \partial_t - \tilde \phi^{-}_{+} - \epsilon^{-}_{+}(\vec k - \tilde {\vec A}^{-}_{+})
	\end{pmatrix} \\
	&= \mathcal{G}_{b} - \chi,
\end{align}
Here, $\mathcal G_b^{-}$ is the Green's function without phase fluctuations and external potentials, and $\chi$ consists of the  phase
derivative and potential terms.  We group these together into effective potentials,
$(\tilde \phi^{\xi}_{\eta}, \tilde{\vec A}^{\xi}_{\eta}) =
(\phi^{\xi}_{\eta} + \partial_t [\theta^{\xi-}_{\eta+1}+\bar\theta], \vec A^{\xi}_{\eta} -  \nabla[\theta^{\xi-1}_{\eta+1}+\bar\theta] ) $.  We then have explicitly
\begin{align}
	\mathcal{G}_b &=
	\begin{pmatrix}
		i \partial_t - \epsilon^{+}_{-}(\vec k ) & d_1  & d_2 & d_3  \\
		d_1 & i \partial_t - \epsilon^{-}_{-}(\vec k ) & d_3  & d_2  \\
		d_2 & d_3 & i \partial_t - \epsilon^{+}_{+}(\vec k ) & d_1 \\
		d_3 & d_2 & d_1  & i \partial_t - \epsilon^{-}_{+}(\vec k )
	\end{pmatrix}, \\
\\
	\chi(q) &= \mathrm{diag}(
	\tilde \phi^{+}_{-}(q) - \tilde{\epsilon}^{+}_{-}(\tilde A^{+}_{-x}(q)),
	\tilde \phi^{-}_{-}(q) - \tilde{\epsilon}^{-}_{-}(\tilde A^{-}_{-1x}(q)),
	\tilde \phi^{+}_{+}(q) - \tilde{\epsilon}^{+}_{+}(\tilde A^{+}_{+x}(q)),
	\tilde \phi^{-}_{+}(q) - \tilde{\epsilon}^{-}_{+}(\tilde A^{-}_{+x}(q))
	),
\end{align}
where $\tilde{\epsilon}(\vec{k}) = \epsilon(\vec{k}) - \epsilon(0) $. The action acquires corrections that can be expressed as a power series in $\chi$, and the Goldstone modes emerge as normal modes of this, in particular from the second order contribution.
In the present context this is given by
\begin{align}
	\label{eq:action}
	S^{(2)}[\{\theta_i\}] &= -\frac{T}{2L^2} \sum_{pq} \mathrm{Tr}[{{\mathcal G}_b}(p) \chi(q) {{\mathcal G}_b}(p+q) \chi(-q)]\\
	&\approx -\frac{T}{2L^2} \sum_{pq} \mathrm{Tr}[{{\mathcal G}_b}(p) \chi(q) {{\mathcal G}_b}(p) \chi(-q)] = -\frac{T}{2L^2} \sum_{pq} \sum_{mn}{{\mathcal G}_b}(p)_{mn}^2 \chi(q)_{nn}\chi(-q)_{mm}.
\end{align}
This may be recast in matrix form,
\newcommand{\gmat}{
	\begin{pmatrix}
		\mathcal G_b(p)^2_{11} &  \mathcal G_b(p)^2_{12} &  \mathcal G_b(p)^2_{13} &  \mathcal G_b(p)^2_{14} \\
		\mathcal G_b(p)^2_{12} &  \mathcal G_b(p)^2_{22} &  \mathcal G_b(p)^2_{23} &  \mathcal G_b(p)^2_{24} \\
		\mathcal G_b(p)^2_{13} &  \mathcal G_b(p)^2_{23} &  \mathcal G_b(p)^2_{33} &  \mathcal G_b(p)^2_{34} \\
		\mathcal G_b(p)^2_{14} &  \mathcal G_b(p)^2_{24} &  \mathcal G_b(p)^2_{34} &  \mathcal G_b(p)^2_{44}
	\end{pmatrix}
}
\newcommand{\jmat}
{
	\begin{pmatrix}
		\mathcal G_b(p)^2_{11} & \mathcal G_b(p)^2_{12} & \mathcal G_b(p)^2_{13} & \mathcal G_b(p)^2_{14} \\
		-\mathcal G_b(p)^2_{12} & -\mathcal G_b(p)^2_{22} & -\mathcal G_b(p)^2_{23} & -\mathcal G_b(p)^2_{24} \\
		\mathcal G_b(p)^2_{13} & \mathcal G_b(p)^2_{23} & \mathcal G_b(p)^2_{33} & \mathcal G_b(p)^2_{34} \\
		-\mathcal G_b(p)^2_{14} & -\mathcal G_b(p)^2_{24} & -\mathcal G_b(p)^2_{34} & -\mathcal G_b(p)^2_{44}
	\end{pmatrix}
}
\begin{align}
	&	S^{(2)}[\{\theta_i\}] = \sum_{q}
	\begin{pmatrix}
		\chi_{11}(-q) & \chi_{22}(-q) & \chi_{33}(-q) & \chi_{44}(-q)
	\end{pmatrix}
	G_2
	\begin{pmatrix}
		\chi_{11}(q) \\ \chi_{22}(q) \\ \chi_{33}(q) \\ \chi_{44}(q)
	\end{pmatrix}\label{eq:action_final}\\
	\text{with}~~&G_2 =  -\frac{T}{2L^2} \sum_p \gmat.
\end{align}
Currents are found by taking functional derivatives of action $S[\{\theta_i\}]$ with respect to $\vec A^{\eta}_\xi$; e.g., $j^{\eta}_{\xi,x}(q) = \frac{\delta S[\{\theta_i\}]}{\delta A^{\eta}_{\xi,x}(q)}$ is the current for the FA indexed by $\eta$ (surface) and $\xi$ (valley).

For simplicity, we consider a model in which the Fermi arcs are straight and parallel, and have dispersions
\begin{align}
	\epsilon^{\eta}_{\xi}(\vec k) = -\xi k_x -\xi \frac{Q}{2}. \label{eq:dispersionSimple}
\end{align}
In this model, by performing the gradient expansion of the action at the first order, $S^{(1)}$ (not shown here) and taking derivatives, one finds
\begin{align}
	j^{(1)+}_{-,x}(q) = - \frac{v_f T}{L^2} \mathcal G_b(q)_{1,1}, &&
	j^{(1)-}_{-,x}(q) =  -\frac{v_f T}{L^2} \mathcal G_b(q)_{2,2}, &&
	j^{(1)+}_{+,x}(q) = \frac{v_f T}{L^2} \mathcal G_b(q)_{3,3}, &&
	j^{(1)-}_{+,x}(q) =  \frac{v_f T}{L^2} \mathcal G_b(q)_{4,4},
\end{align}
which are persistent currents, independent of the gradients of
the phases.  These are present due to the broken time-reversal symmetry of the WSM's.  Note that they do not imply any net {\it electric} current on the surfaces in equilibrium, as the sums over flavors on each surface are zero.  Deviations from the persistent currents arise due to $S^{(2)}$ and yield contributions proportional to the phase gradients.  These may be
expressed as
\begin{align}
	\begin{pmatrix}
		j^{(2)+}_{-,x}(q) \\
		j^{(2)-}_{-,x}(q) \\
		j^{(2)+}_{+,x}(q) \\
		j^{(2)-}_{+,x}(q) \\
	\end{pmatrix} &=
	D G_2
	\begin{pmatrix}
		\chi_{11}(q) \\ \chi_{22}(q) \\ \chi_{33}(q) \\ \chi_{44}(q)
	\end{pmatrix} =  DG_2	Y(q) U
	\begin{pmatrix}
		\theta_1(q) \\
		\theta_2(q) \\
		\theta_3(q) \\
	\end{pmatrix},
	\label{eq:currentEqn}
\end{align}
with $ D = \mathrm{diag}\left(v_F, v_F, -v_F,-v_F \right)$, $Y(q) =
\mathrm{diag}(
-i (q_0 + v_Fq_x),
-i (q_0 +  v_Fq_x),
-i (q_0 - v_Fq_x),
-i (q_0 - v_Fq_x)
)$, and
$$U =
\frac{1}{2}
\begin{pmatrix}
	1 & 1 & 1\\
	1 & -1 & -1\\
	-1 & -1 & 1\\
	-1 & 1 & -1
\end{pmatrix}.$$
\vspace{.2cm}
The action may be written in terms of the three independent phases in the
form
\begin{align}
	S^{(2)}[{\theta_i}] &=
	\sum_q
	\begin{pmatrix}
		\theta_1(-q) & \theta_2(-q) & \theta_3(-q)	
	\end{pmatrix}
	U^T X(-q)^T G_2 X(q) U
	\begin{pmatrix}
		\theta_1(q) \\
		\theta_2(q) \\
		\theta_3(q)
	\end{pmatrix} 
\end{align}
$U^T X(-q)^T G_2 X(q) U$  is the matrix valued polarizability function, $\Pi(q)$, as mentioned in the main text. By diagonalizing the polarizability function, we find
\begin{align}
	S^{(2)}[{\theta_i}] 	&= \sum_q
	\begin{pmatrix}
		\tilde\theta_1(-q) & \tilde\theta_2(-q) & \tilde\theta_3(-q)	
	\end{pmatrix}
	R(q)
	\begin{pmatrix}
		\tilde\theta_1(q) \\
		\tilde\theta_2(q) \\
		\tilde\theta_3(q)
	\end{pmatrix},
\end{align}
The last line of the above equation employs eigenmodes of the matrix $ U^T Y(-q)^T G_2 Y(q) U  \equiv W R(q) W^{-1}$, where $R(q)$ is a diagonal matrix, and
$(\theta_1,\theta_2,\theta_3)^T = W(\tilde\theta_1,\tilde\theta_2,\tilde\theta_3)^T$.  This last form represents the quadratic action of the Goldstone modes, expressed in terms of the normal mode amplitudes of the second order action.
The flavor currents can then be written
\begin{align}
	\label{eq:super_current}
	\begin{pmatrix}
		j^{(2)+}_{-,x}(q) \\
		j^{(2)-}_{-,x}(q) \\
		j^{(2)+}_{+,x}(q) \\
		j^{(2)-}_{+,x}(q) \\
	\end{pmatrix} &=
	D G_2 Y(q) U W
	\begin{pmatrix}
		\tilde \theta_1(q) \\
		\tilde \theta_2(q) \\
		\tilde \theta_3(q)
	\end{pmatrix}.
\end{align}
In the static limit, we have
$Y(q) =
-i v_F q_x\mathrm{diag}(1,1,-1,-1) \equiv -i v_F q_x \rho_3$. In real space, the currents are written as $$	(j_{\xi}^{\eta})_{x} = \sum_{i=1,2,3}\Gamma_{\eta \xi i}^{x}\nabla_x\tilde{\theta}_i,$$
where $\Gamma^{x} = v_FD G_2 \rho_3 U W$. 

\subsubsection*{Counterflow current}
From the above model we can derive an expression for the counterflow current, and moreover show that there is no configuration of phase gradients that yields a net electric (i.e., ``co-flow'') current at the interface.  This turns out to be most directly a consequence of mirror symmetry of the model about the $k_y$ axis, encoded in the Green's function by the symmetry
\begin{align}
	\mathcal P \mathcal G(p_0,p_x) \mathcal P^{-1} = \mathcal G(p_0,-p_x).\label{eq:symm}
\end{align}
where $\mathcal P = \sigma_x \otimes \mathbb I_2$.
Together with
the dispersions in Eq.~\eqref{eq:dispersionSimple}, this symmetry results in a total
cancellation of currents, i.e., $\sum_{\eta\xi}(j^{\eta}_{\xi})_x = 0$, which can be seen as follows.  Writing Eq.~(\ref{eq:currentEqn}) in the static limit, ($q_0 \to 0$), one finds
\begin{align}
	\begin{pmatrix}
		j^{(2)+}_{-,x}(q) \\
		j^{(2)-}_{-,x}(q) \\
		j^{(2)+}_{+,x}(q) \\
		j^{(2)-}_{+,x}(q) \\
	\end{pmatrix} &=
	- iq_x v_F^2 \frac{T}{2L^2} \sum_p \begin{pmatrix}
		-	\mathcal G_b(p)^2_{11} &  -\mathcal G_b(p)^2_{12} &  \mathcal G_b(p)^2_{13} &  \mathcal G_b(p)^2_{14} \\
		-\mathcal G_b(p)^2_{12} &  -\mathcal G_b(p)^2_{22} &  \mathcal G_b(p)^2_{23} &  \mathcal G_b(p)^2_{24} \\
		\mathcal G_b(p)^2_{13} &  \mathcal G_b(p)^2_{23} &  -\mathcal G_b(p)^2_{33} &  -\mathcal G_b(p)^2_{34} \\
		\mathcal G_b(p)^2_{14} &  \mathcal G_b(p)^2_{24} &  -\mathcal G_b(p)^2_{34} &  -\mathcal G_b(p)^2_{44}
	\end{pmatrix}U
	\begin{pmatrix}
		\theta_1(q) \\
		\theta_2(q) \\
		\theta_3(q) \\
	\end{pmatrix}.
\end{align}
The symmetry Eq.~(\ref{eq:symm}) implies
\begin{align}
	-\frac{T}{2L^2} \sum_p \begin{pmatrix}
		-	\mathcal G_b(p)^2_{11} &  -\mathcal G_b(p)^2_{12} &  \mathcal G_b(p)^2_{13} &  \mathcal G_b(p)^2_{14} \\
		-\mathcal G_b(p)^2_{12} &  -\mathcal G_b(p)^2_{22} &  \mathcal G_b(p)^2_{23} &  \mathcal G_b(p)^2_{24} \\
		\mathcal G_b(p)^2_{13} &  \mathcal G_b(p)^2_{23} &  -\mathcal G_b(p)^2_{33} &  -\mathcal G_b(p)^2_{34} \\
		\mathcal G_b(p)^2_{14} &  \mathcal G_b(p)^2_{24} &  -\mathcal G_b(p)^2_{34} &  -\mathcal G_b(p)^2_{44}
	\end{pmatrix} = -r_1 \sigma_0\otimes\sigma_0-r_2 \sigma_0\otimes\sigma_x + r_3 \sigma_x\otimes\sigma_0 + r_4 \sigma_x\otimes\sigma_x,
\end{align}
where $r_i = -\frac{T}{2L^2} \sum_p \mathcal{G}_p(p)^2_{1i}$. It is easy to verify, using the form of $U$, that $\sum_{s_1,s_2} j_{s_1,x}^{s_2}=0$. Furthermore, one obtains, in real space,
\begin{align}
	j^{+}_{-,x}+j^{+}_{+,x} = -(j^{-}_{-,x}+j^{-}_{+,x}) = 2 v_F^2 \partial_x\theta_3 (-r_1-r_2+r_3+r_4),\label{eq:cfc}
\end{align}
with $\theta_3 = (\thad - \thbd + \thcd) /2$.  This implies
if (dissipationless) current is driven on one of the surfaces ($\sum_{\xi}j^{\eta}_{\xi}$), then due
to interlayer coherence an equal and opposite current flows in the other
($\sum_{\xi}j^{\bar\eta}_{\xi}$). We note that this is a consequence of the inter-locking phase relations Eq.~(\ref{eq:phase1}).
We note that, in case some of the orders vanish, the original phase relations, Eq.~(\ref{eq:phase1}) do not hold, and subsequently the restriction  on the counter-flow current,  Eq.~(\ref{eq:cfc}), may not hold anymore.

\section{Three Fermi Arcs}%
\label{app:3_fermi_arcs}
In this section we provide more details for the equations of the interacting Green's function that we solve numerically for the three FA system. The FAs are in a hexagonal Brillouin-zone, and we consider two orientations of the surfaces, as shown in Fig.~\ref{fig:3fa_data} (a) and (e). The construction of these FAs broadly follows that of Appendix~~\ref{app:construction_of_2_pair_of_tilted_fermi_arcs} and Appendix~~\ref{app:interacting_hamiltonian}, where the FAs are indexed by the surface index $\eta = \pm$, and valley index $ \xi \in \{1,2,3\}$.

\subsubsection*{Exciton major configuration}
\begin{figure}[t]
	\centering
	\includegraphics[width=0.95\textwidth]{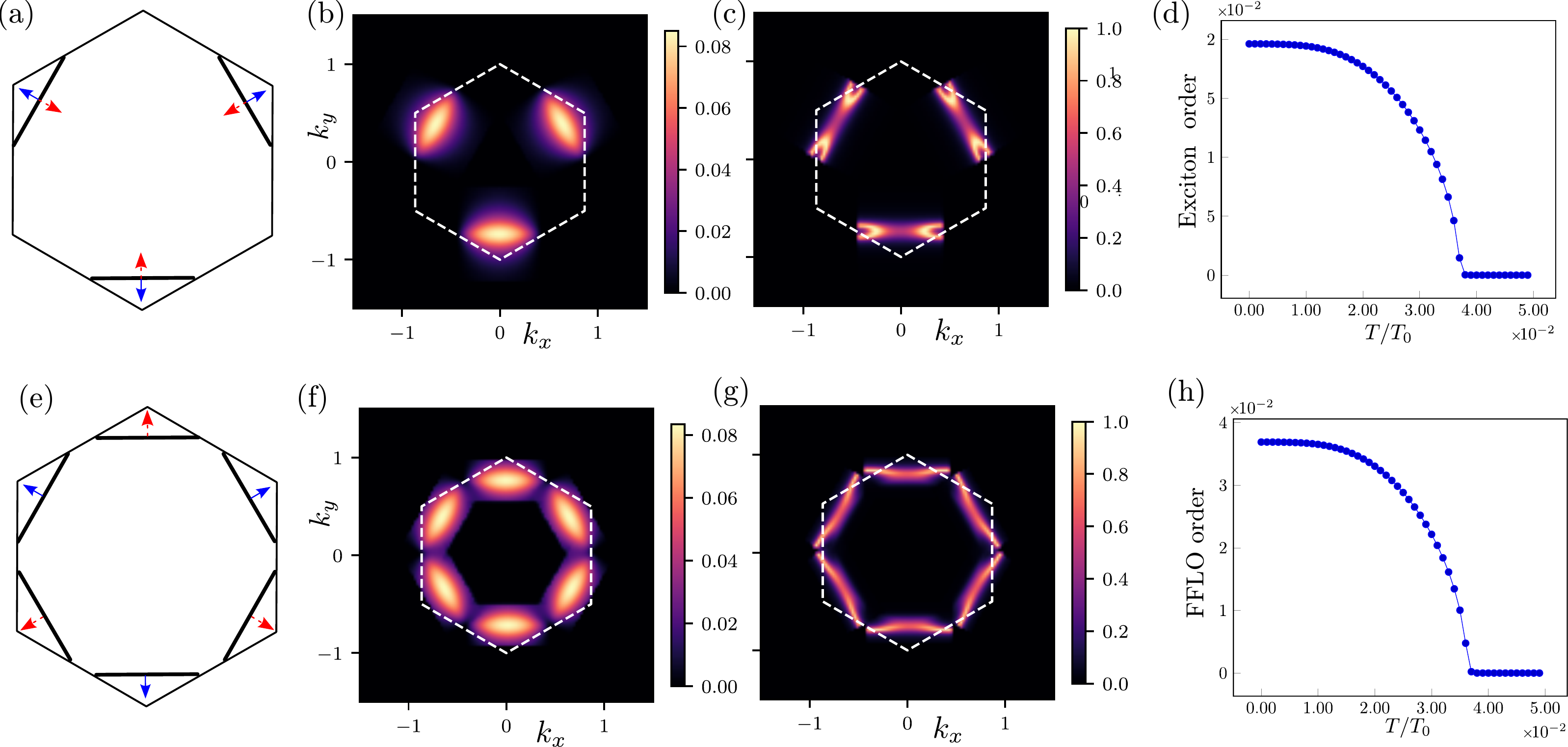}
	\caption{
	The case of three Fermi-arcs systems, where we consider three linear Fermi-arcs in a hexagonal Brillouin zone, with two possible orientations between the surfaces, shown in (a) and (e). The blue lines show the Fermi arcs on the Brillouin zone of one surface and the red lines show the Fermi arcs on that of the second surface. (b), (c), (d) shows, respectively, the excitonic order, the spectral function ($ A(\vec k, \omega) = -2 {\rm Im}\mathcal G(\vec k, \omega + i 0^+)$ at $\omega=0.6$) and the second order transition as a function of temperature, for the orientation in (a). Analogous plots are shown for the orientation (e), in (f), (g) and (h). The length of the FAs are 0.18\AA$^{-1}$. The momentum cutoff for individual FA's is determined using an energy cutoff of $|\epsilon_{\rm cut} | \approx 0.22$.  The separation of the surfaces is assumed to be $t=1$. All of the above results are obtained self-consistently to a tolerance of $10^{-6}$, with a momentum grid of 1795 points for numerical integration.  }
	\label{fig:3fa_data}
\end{figure}

This is the configuration of Fig.~\ref{fig:3fa_data} (a). As there is no finite-momentum nesting among the three FAs, CDW and FFLO-exciton orders are not energetically favorable.  Accordingly we neglect these orders, and consider only excitonic self-energy terms for the numerical analysis. All the self-energies in this case are given by (a) and (d) of Fig.~\ref{fig:SE}.  Since the modes are excitonic, we use the notation
$\Sigma^{+-}_{\eta\eta} \equiv \Sigma_{\eta}$, $\mathcal G^{+-}_{\xi\xi}\equiv G_{\xi}$ and $V^{-+}_{\xi\xi\xi\xi}
\equiv V_{\xi}$. The 6$\times$6 matrix valued Green's function is given
by $ {\mathcal G} = [({\mathcal G}^{0})^{-}-{\Sigma}]^{-}$, where
\begin{align}
	\Sigma_{\xi}(\vec k) =
	-\frac{1}{\beta} \sum\limits_{m,\vec q}  V_{\xi}(\vec k, \vec k, \vec q) \mathcal G_{\xi}(\vec k - \vec q, \omega_n - \omega_m)  e^{i(\omega_n - \omega_m) 0^{+}} + \mathcal O\left(\frac{e^{-k_0 t}}{k_0}\right),
\end{align}
which gives 
\begin{align}
	&\mathcal G_{\xi}(\vec k, \omega_n) = \frac{\Sigma_{\xi}(\vec k)}{-\omega_n^2 - \mathcal D^2(\vec k, \omega_n)},~\mathcal D^2(\vec k, \omega_n) = \left({\epsilon^{+}_{\xi}}(\vec k)\right)^2 + \abs{\Sigma_{\xi}(\vec k)}^2.\\
\end{align}
After performing the Matsubara summation, one obtains
\begin{align}
	\Sigma_{\xi}(\vec k) = \frac{1}{2}\sum\limits_{\vec q}
		V_{\xi}(\vec k,\vec k,\vec q) \Sigma_{\xi}(\vec k - \vec q)  \frac{ \tanh(\beta \mathcal D(\vec k-\vec q) /2)}{ \mathcal D(\vec k-\vec q)}  + \mathcal O\left(\frac{e^{-k_0 t}}{k_0}\right)
	.
\end{align}
We present numerical results for the self-energies, spectral functions $A(\vec k, \omega) = -2 {\rm Im}\mathcal G(\vec k, \omega + i 0^+)$, and the variation of the self-energy with temperature (indicating a continuous phase transition at the onset of excitonic order) in Fig.~\ref{fig:3fa_data}.

\subsubsection*{FFLO major configuration}
If one of the WSM surfaces with three FA's is rotated by
$60^{\circ}$ relative to the other, the FAs in the opposite layer become nested
pairwise (see Fig.~\ref{fig:3fa_data} (e)), leading to formation of the FFLO-exciton orders. The self-energy diagrams for this case are same as (d) of Fig.~\ref{fig:SE}. We proceed exactly in the same way as the last section to compute the FFLO self-energies, and obtain
\begin{align}
	&\Sigma^{+-}_{\xi\bar\xi}(\vec k, \vec k - \vec Q_{\xi\bar\xi}) = \frac{1}{2}\sum_{\vec q}
\Sigma^{+-}_{\xi\bar\xi}(\vec k - \vec q, \vec k - \vec q - \vec Q_{\xi\bar\xi}) V^{+-}_{\xi\xi\bar\xi\bar\xi}(\vec k, \vec k - \vec Q_{\xi\bar\xi},\vec q) \frac{ \tanh(\beta \mathcal D(\vec k-\vec q) /2)}{ \mathcal D(\vec k-\vec q)},
\end{align}
where $\mathcal D(\vec k)^2 = \left(\epsilon^{+}_{\xi}(\vec k)\right)^2 + \abs{\Sigma^{+}_{\xi\bar\xi}(\vec k, \vec k - \vec Q_{\xi\bar\xi})}^2$. Numerical results for the self-energies, spectral density of states and the variation with temperature are presented in Fig.~\ref{fig:3fa_data}.

\twocolumngrid

\end{document}